# Control of Goos-Hänchen shift of a light beam via a coherent driving field

Li-Gang Wang, [1, 2] Manzoor Ikram,[1] and M. Suhail Zubairy[1, 3]

[1]*Centre for Quantum Physics, COMSATS Institute of Information Technology, Islamabad, Pakistan*

[2]*Department of Physics, Zhejiang University, Hangzhou 310027, China*

[3] *Institute for Quantum Studies and Department of Physics, Texas A&M University, College Station, Texas 77843, USA and Texas A&M University at Qatar, P. O. Box 23874, Education City, Doha, Qatar*

Abstract

We present a proposal to manipulate the Goos-Hänchen shift of a light beam via a coherent control field, which is injected into a cavity configuration containing the two-level atomic medium. It is found that the lateral shifts of the reflected and transmitted probe beams can be easily controlled by adjusting the intensity and detuning of the control field. Using this scheme, the lateral shift at the fixed incident angle can be enhanced (positive or negative) under the suitable conditions on the control field, without changing the structure of the cavity.

It is well known that there exists a tiny lateral shift between the totally reflected light beam and the incident light beam when a total reflection happens at the interface of two media [1]. This lateral shift is known as Goos-Hänchen shift [2-3] and it is usually proportional to the penetration depth with an order of a wavelength. The Goos-Hänchen shift has potential applications in various fields such as micro-/nano-optics, acoustics, quantum mechanics and plasma physics [4]. In the last two decades, various structures containing different kinds of media were analyzed in order to achieve the enhanced (positive or negative) lateral shift [5-18], such as, near the angle of the Brewster dip on reflection from a weakly absorbing semi-infinite medium [6-7], near the angle of the edge inside the photonic bandgap [9] and the angle of defect mode of photonic crystals [10], and in the slab systems containing different artificial medium [11]. In 2003, Li [12] found that the lateral shift of a light beam passing through a lossless dielectric slab can be negative when the incident angle is larger than the threshold angle. It was further shown that the lateral shift of the reflected beam can be greatly negative near the resonance in a weakly absorbing dielectric slab [13]. Recently, Yan et. al. [14] reported that the lateral shift of a light beam passing through a gain medium can also be large (positive or negative) near the resonance or the critical angle of the slab.

However, in all these investigations, the lateral shift cannot be manipulated for a fixed configuration or device. The situation changes if we have some atomic medium in a fixed cavity, knowing that the dispersion-absorption relations can be dramatically modified by using coherent driving fields in the atomic systems [15]. In 1991, Scully [16] presented the idea of the enhancement of refractive index with vanishing absorption, by preparing the lower level doublet of the atom in a coherent superposition. Following this suggestion, many others [17] proposed various schemes to manipulate the refractive index and absorption in different systems. The driven two-level systems have been investigated in details. For example, Wilson-Gordon et. al. [18] and Quang et. al. [19] showed the modification of the dispersion-absorption relation in two-level atomic systems by a strong coherent external field. In 1993, Pfleghaar et. al. [8] observed the transition of the lateral shift of a probe light beam totally reflected from a glass-cesium vapor interface near the critical angle by changing the detuning of the probe beam.

In this letter, we propose a scheme to control the lateral shift of a probe beam reflected from or transmitted through a cavity containing two-level atomic medium, via an external coherent control field. By modifying the dispersion-absorption properties of the intracavity medium, the resonant conditions of the slab or cavity system are expected to be changed dramatically. Therefore it is expected that the lateral shifts of the reflected and transmitted probe beams can be easily controlled by adjusting the intensity and detuning of the external control field.

We consider a TE-polarized probe light beam $E_p$ with angle $\theta$ incident from the vacuum ($\varepsilon_0 = 1$) upon the cavity containing a two-level atomic medium ($\varepsilon_2$), as shown in Fig. 1(a). The cavity is composed of two nonmagnetic dielectric slabs ($\varepsilon_1$) with identical thickness $d_1$, and the intracavity two-level atomic medium is with thickness $d_2$. For the sake of simplicity, we assume that the strong coherent control beam is homogenously normal incident on the interface of the cavity and the frequency of the control beam is far away from the resonant modes of the cavity, so that the control beam (corresponding to the Rabi frequency $\Omega_c$) could be assumed to be homogenously applied onto the transition between the the excited state $|a\rangle$ and the ground state $|b\rangle$, as shown in Fig. 1(b). In an appropriate frame and under the dipole approximation, the system's Hamiltonian is written in the form

$$H_I = \hbar(\Delta_c + \Delta_p)\sigma_{aa} - \frac{\hbar}{2}(\Omega_c \sigma_{ab} + \Omega_c^* \sigma_{ba}) - \frac{\hbar}{2}(\Omega_p \sigma_{ab} + \Omega_p^* \sigma_{ba}), \tag{1}$$

with detunings $\Delta_c = \omega_{ab} - \omega_c$, $\Delta_p = \omega_{ab} - \omega_p$, $\delta \equiv \Delta_c - \Delta_p = \omega_p - \omega_c$, $\omega_{c,p}$ are the frequencies of the control and probe beams, and $\omega_{ab}$ is the transition frequency from $|a\rangle$ to $|b\rangle$. $\sigma_{ij} = |i\rangle\langle j|$ ($i, j = a, b$) are population operators for $i = j$ and dipole operators for $i \neq j$. From the master equation [15], the linear susceptibility $\chi(\omega_p)$ of the two-level atomic medium for the probe beam is given by [19]

$$\chi = \beta \frac{(2\Delta_c - i\gamma)\{(\delta + i\gamma)(\gamma - i2\Delta_c)[2(\delta + \Delta_c) + i\gamma] - i2\delta|\Omega_c|^2\}}{(4\Delta_c^2 + \gamma^2 + 2|\Omega_c|^2)\{(\gamma - i\delta)[4\Delta_c^2 - (2\delta + i\gamma)^2] + 2|\Omega_c|^2(\gamma - i2\delta)\}}, \tag{2}$$

where $\beta = 2N|u_{ab}|^2/\varepsilon_0\hbar$, $u_{ab}$ is the electric dipole moment of the transition $|a\rangle \leftrightarrow |b\rangle$, $N$ is the atomic density, and $\gamma$ denotes the atom decays from $|a\rangle$ to $|b\rangle$. From the relation $\varepsilon_2 = \sqrt{1 + \chi}$, it is obvious that the dielectric function of the intracavity medium is frequency-dependent and it can be controlled by the external control field. It should be pointed out that the first experiment by Zibrov et. al. [20] confirmed the resonant enhancement of the index of refraction reaching $\Delta n \approx 10^{-4}$. Recently Zibrov et. al. [21] further observed the maximum resonant change in the refractive index with the $\Delta n \approx 0.1$. Based on the measured change of the index of refraction, for simplicity, we take the reasonable parameters: $\gamma = 1 \text{MHz}$ and $\beta = 2\gamma$ in all the following theoretical and numerical calculations. Under these typical parameters the refractive index of the intracavity medium can change within a range of $\Delta n \approx 10^{-1} - 10^{-2}$ and has the absorption or gain property by changing the external control field.

In our configuration, we can apply the standard characteristic matrix approach [22-23] to calculate the amplitude transmission $T(k_y, \omega_p)$ and reflection $R(k_y, \omega_p)$ for the weak probe beam through the cavity at a given probe frequency $\omega_p$. The transfer matrix of the $j$ th layer can

be expressed as [23]

$$M_j(k_y, \omega_p, d_j) = \begin{pmatrix} \cos[k_z^j d_j] & i\sin[k_z^j d_j]/q_j \\ iq_j \sin[k_z^j d_j] & \cos[k_z^j d_j] \end{pmatrix}, \quad (3)$$

where $k_z^j = \sqrt{\varepsilon_j k^2 - k_y^2}$ is the $z$ component of the wavenumber in the $j$ th layer, $q_j = k_z^j/k$, $d_j$ is the thickness of the $j$ th layer, $k_y$ is the $y$ component of the wavenumber ($k = \omega_p/c$) in vacuum, and $c$ is the light speed in vacuum. The total transfer matrix for the considering cavity is given by $Q(k_y, \omega_p) = M_1(k_y, \omega_p, d_1) M_2(k_y, \omega_p, d_2) M_3(k_y, \omega_p, d_1)$. Therefore, the coefficients $R$ and $T$ are, respectively, given by

$$R(k_y, \omega_p) = \frac{q_0(Q_{22} - Q_{11}) - (q_0^2 Q_{12} - Q_{21})}{q_0(Q_{22} + Q_{11}) - (q_0^2 Q_{12} + Q_{21})}, \quad (4)$$

$$T(k_y, \omega_p) = \frac{2q_0}{q_0(Q_{22} + Q_{11}) - (q_0^2 Q_{12} + Q_{21})}, \quad (5)$$

where $q_0 = k_z/k$ (here $k_z$ is the z component of the wavenumber in vacuum), and $Q_{ij}$ are the elements of the matrix $Q(k_y, \omega_p)$. We can make similar analysis for the TM-polarized probe beam.

For a well-collimated probe beam with a sufficiently large width (i.e., with a narrow angular spectrum, $\Delta k \ll k$), according to the stationary phase theory [12, 24], the lateral shifts of the reflected and transmitted probe beams can be calculated analytically using $S_{r,t} = -(\lambda/2\pi)[d\phi_{r,t}/dk_y]$, where $\phi_{r,t}$ are the phases of $R$ and $T$. Following the method in the reference [13], the lateral shifts of both the reflected and transmitted probe beams can be expressed as:

$$S_r = -\frac{\lambda}{2\pi} \frac{d\phi_r}{dk_y}$$
$$= -\frac{\lambda}{2\pi} \frac{1}{|R|^2} \left\{ \text{Re}[R] \frac{d\,\text{Im}[R]}{dk_y} - \text{Im}[R] \frac{d\,\text{Re}[R]}{dk_y} \right\}, \quad (6)$$

$$S_t = -\frac{\lambda}{2\pi} \frac{d\phi_t}{dk_y}$$
$$= -\frac{\lambda}{2\pi} \frac{1}{|T|^2} \left\{ \text{Re}[T] \frac{d\,\text{Im}[T]}{dk_y} - \text{Im}[T] \frac{d\,\text{Re}[T]}{dk_y} \right\}. \tag{7}$$

From Eqs. (6) and (7), we can calculate the dependence of the lateral shifts of the reflected and transmitted probe beams on the external control field. In the following, we take the fixed parameters for the considering cavity as follows: $d_1 = 0.2\,\mu\text{m}$, $d_2 = 5\,\mu\text{m}$, and $\varepsilon_1 = 2.22$.

Now we discuss in detail the effects of the intensity and the detuning of the control beam on the lateral shifts. First we set the detuning of the probe beam $\Delta_p = -5\gamma$. Figure 2 shows the dependence of the reflected and transmitted lateral shifts ($S_r$ and $S_t$) on the incident angle of the probe beam under different strengths of the control beam. It is clear seen that the lateral shifts for the reflected and transmitted probe beams changes dramatically when the control field ($\Omega_c$) changes. In Figs. 2(a), 2(b) and 2(c), the reflected probe beam suffers the large negative shift near the resonant condition of the cavity, while the transmitted probe beam suffers the positive shift. The dependence of the lateral shifts on the control beam strength is due to the fact that the linear susceptibility $\chi(\omega_p)$ of the intracavity medium varies with the change of $\Omega_c$. This variation in $\chi(\omega_p)$ with $\Omega_c$ modifies the resonant condition of the cavity and we observe the manipulation effect of the control beam on the lateral shifts. In Ref. [13], it has been shown that there exists a large negative shift near resonance due to the weak absorption of the dielectric slab. However, as in many previous proposals [9-14], the lateral shifts of the reflected and transmitted beams can not be manipulated once one chooses the structure. Here we use the external control light field to manipulate the susceptibility of the intracavity medium, therefore the lateral shifts are changed as the control light field changes. When the value of $\Omega_c$ is further increased, it is found that both $S_r$ and $S_t$ can be enhanced and can be positive or negative at different angles as shown in Fig. 2(d-f). In fact, the susceptibility in these cases becomes gain, which also leads to large lateral shifts (positive or negative) near resonances [14]. *Thus it is a very useful and powerful way to control the susceptibility in order to manipulate the lateral shifts by using the external control field.*

In Fig, 3, we plot the dependence of the lateral shifts on the control field under different incident angles. Here we see that the lateral shifts for both the reflected and transmitted probe beam are strongly dependent on the value of $\Omega_c$, and at different incident angle the lateral shifts have different behaviors on the control field. The lateral shifts can be very large (negative or positive) at certain value of $\Omega_c$ at the fixed incident angle of the probe beam. Using this effect, we can easily manipulate the lateral shifts of the reflected and transmitted probe beams without changing or adjusting the structure of the medium.

Now we consider the effect of another control parameter, the detuning $\Delta_c$ of the control beam on the lateral shifts. Here we set $\Delta_p = 5\gamma$ and then change $\Delta_c$ with constant $\Omega_c$. It is shown in Fig. 4 that the lateral shifts of the reflected and transmitted probe beams also depend on the detuning $\Delta_c$ of the control field. It is due to the fact that, the susceptibility of the intracavity medium is modified as $\Delta_c$ changes, which leads to the modification of the resonant condition of the cavity. Therefore the enhancement or suppression effect of the lateral shift is observed by changing the control parameters $\Delta_c$ and $\Omega_c$ of the coherent field.

In the above discussions, the results are based on the stationary-phase theory [24], in which the incident probe beam is assumed to be the plane wave. In the following discussion, we will show that our results are still valid in the real system when the incident probe beam has a finite width (typically, with a Gaussian profile). Here we emphasize again that the probe beam is very weak so that the linear susceptibility of the intracavity medium is always valid. Following the method described in the reference [10, 25], we can easily simulate the propagation of the weak Gaussian-shaped probe beam passing through the cavity containing the two-level atomic medium. The electric field of the incident probe beam at the plane of $z=0$ can be given by the following integral form: $E_x^{(i)}(y)\big|_{z=0} = (1/2\pi)^{1/2} \int A(k_y) \exp(ik_y y) dk_y$, where $A(k_y) = \frac{W_y}{\sqrt{2}} \exp[-\frac{W_y^2 (k_y - k_{y0})^2}{4}]$ is the initial angular spectrum distribution of the Gaussian-shaped probe beam with an incident angle $\theta$, $k_{y0} = k\sin\theta$, $W_y = W/\cos\theta$, and $W$ is the half-width of the probe beam at the incident plane of $z=0$. Using the amplitude reflection and transmission coefficients [Eqs. (4-5)], the electric fields of the reflected and transmitted probe beams can be written as [10, 25]: $E_x^{(r)}(y)\big|_{z=0} = (1/2\pi)^{1/2} \int R(k_y, \omega_p) A(k_y) \exp(ik_y y) dk_y$, and $E_x^{(t)}(y)\big|_{z=2d_1+d_2} = (1/2\pi)^{1/2} \int T(k_y, \omega_p) A(k_y) \exp(ik_y y) dk_y$. In our simulations, we take $W = 100\lambda_p$ and $W = 600\lambda_p$, which are substantially larger than the wavelength $\lambda_p$ of the probe beam, so that $A(k_y)$ is sharply distributed around $k_{y0}$. Figure 5 shows the normalized intensity profiles of the reflected and transmitted probe beams for two different cases: one is for the large negative lateral shifts when $\Omega_c = 8\gamma$, and another is for the large positive lateral shifts when $\Omega_c = 8.45\gamma$. It is shown that the lateral shifts of the reflected and transmitted probe beams can be controlled to be negative or positive with the adjustment of the control field. From Fig. 5, it is also clear that the shapes of the reflected and transmitted probe beams are nearly the same as the incident field for the case of the probe beams with a larger width.

In summary, we have proposed a scheme to realize the manipulation on the lateral shift of a light probe beam via a coherent control field, which is applied onto the two-level atoms inside a cavity. By adjusting the parameters of the external coherent driving field, the susceptibility of the intracaivity medium can be modified, which changes the resonance condition of the cavity. Therefore it is found that the lateral shifts of the reflected and transmitted probe beams can be controlled by adjusting the intensity and detuning of the control field. This proposal provides a convenient and powerful tool for controlling the lateral shift of the probe beam without changing the structure of the cavity, and it also provides a possibility for obtaining the large negative or positive lateral shift by changing the external controlling field. Finally the numerical simulation gives a solid confidence in our theoretical predication using the stationary phase method.

The authors thank COMSTECH for its support. LG.W gratefully acknowledges the hospitality at Centre for Quantum Physics, Pakistan, where this work was done and the support from NSFC (No.10604047) and Zhejiang Province Education Foundation (No.G20630). MSZ would like to thank Alexander von Humboldt Foundation for a senior Humboldt Research Award.

# Figure Captions

FIG. 1 (Color online) (a) Schematic of a cavity containing a two-level atomic medium, and (b) the energy scheme of the two-level atom. The blue and red arrows denote the coherent controlling field and the probe fields, respectively; and $S_r$ and $S_t$ denote the lateral shifts of the reflected and transmitted probe beams.

Fig. 2. (Color online) The dependence of both $S_r$ (solid) and $S_t$ (dashed) on the incident angle under different controlling Rabi frequencies: (a) $\Omega_c = 0$, (b) $\Omega_c = 3\gamma$, (c) $\Omega_c = 4.95\gamma$, (d) $\Omega_c = 5.5\gamma$, (e) $\Omega_c = 6.5\gamma$, and (f) $\Omega_c = 7.5\gamma$, with $\omega_{ab} = 2\pi \times 300$ THz, $\Delta_c = 0$, $\beta = 2$ MHz, $\gamma = 1$ MHz, and $\Delta_p = -5\gamma$.

Fig. 3. (Color online) The dependence of both $S_r$ (solid) and $S_t$ (dashed) on the controlling Rabi frequency under different incident angles: (a) $\theta = 30°$ and (b) $\theta = 60°$. Other parameters are the same as in Fig. 2.

Fig. 4. (Color online) The dependence of both $S_r$ (solid) and $S_t$ (dashed) on the detuning $\Delta_c$ at different incident angles: (a) $\theta = 30°$ and (b) $\theta = 60°$, with $\Omega_c = 6\gamma$ and $\Delta_p = 5\gamma$. Inset in (b) shows the lateral shifts for certain values of $\Delta_c$. Other parameters are the same as in Fig. 2.

FIG. 5. Numerical simulations of the reflected and transmitted beams from the cavity under different Rabi frequencies: (a) and (b) $\Omega_c = 8\gamma$, and (c) and (d) $\Omega_c = 8.45\gamma$ for (a) and (c) $W = 100\lambda_p$ and for (b) and (d) $W = 600\lambda_p$, at $\theta = 60°$. The dashed, thin-solid, and thick-solid curves denote the incident, reflected and transmitted probe beams, respectively. Other parameters are the same as in Fig. 2.

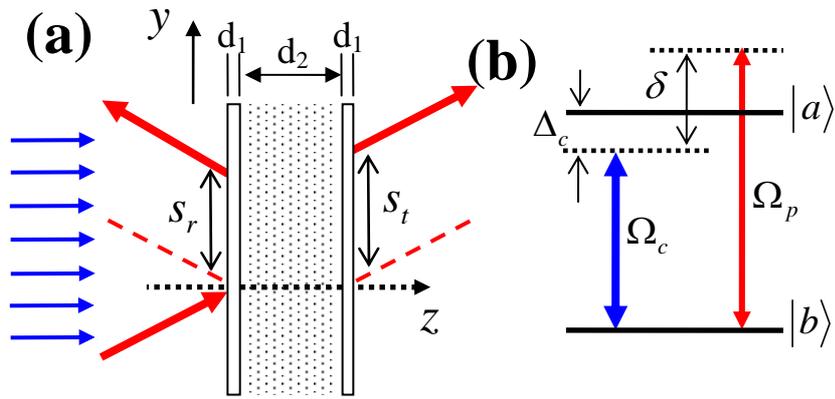

FIG. 1

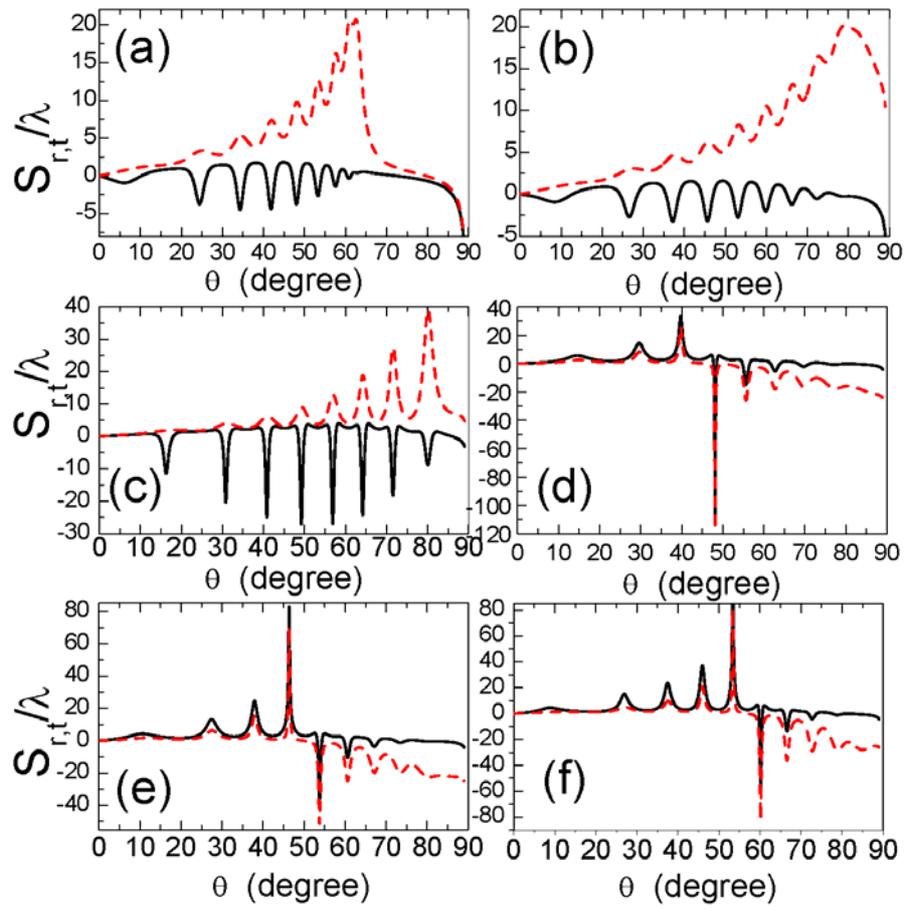

FIG. 2

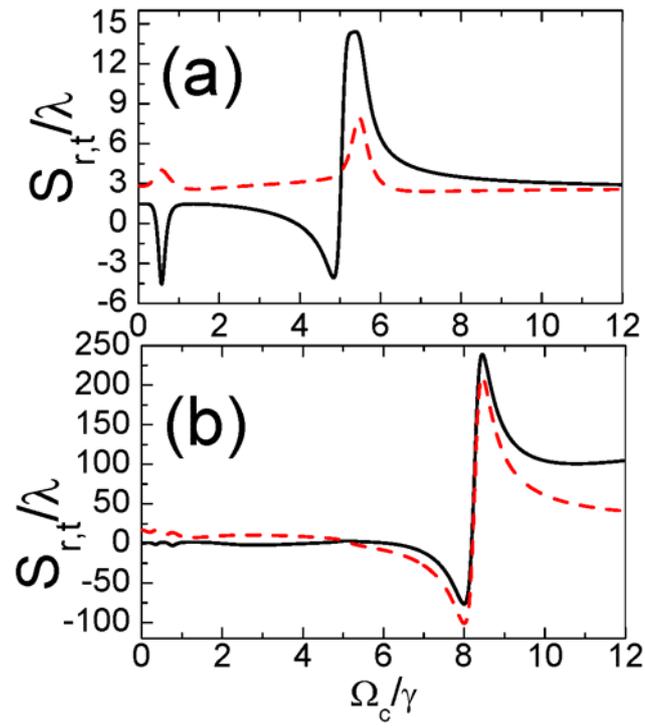

**FIG. 3**

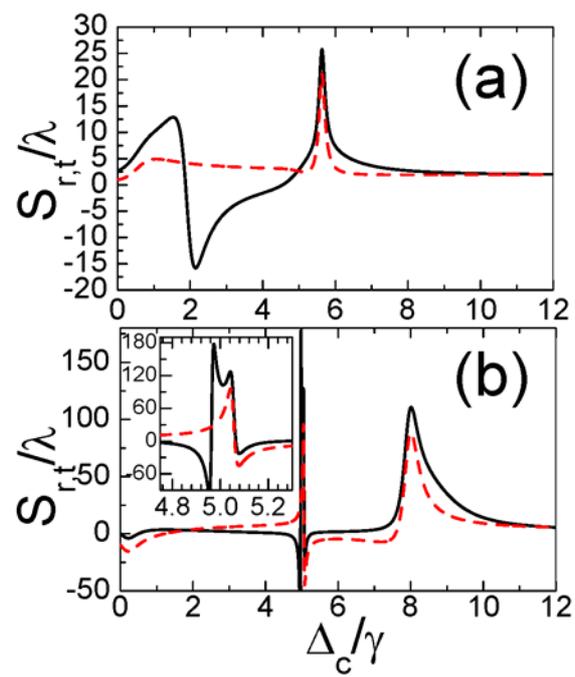

**FIG. 4**

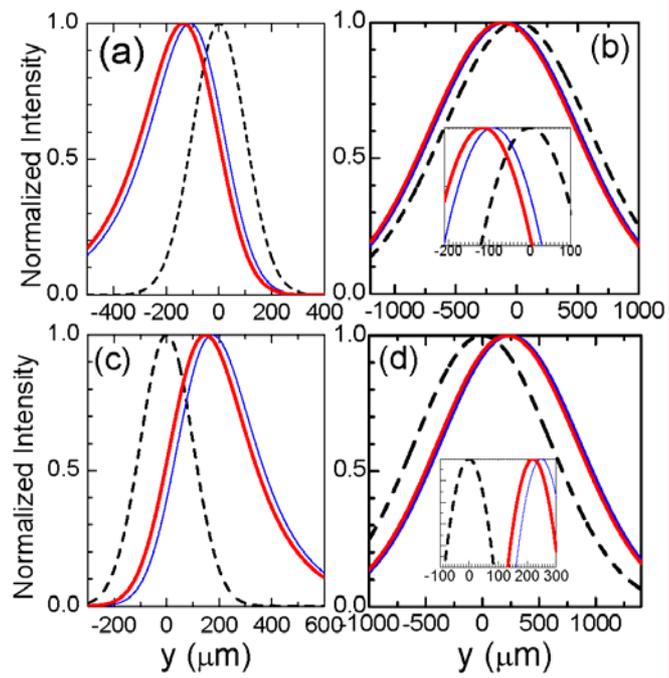

**FIG. 5**